\documentclass[11pt,twoside]{article}

\usepackage{asp2006}
\usepackage{lscape}
\usepackage{graphicx}
\usepackage{commands_def}
\usepackage{natbib} \bibpunct{(}{)}{;}{a}{}{,} 
\begin{document}
\title{Disk-Halo interface: The ``foot point'' of the Galactic Molecular Loops}   
\author{D. Riquelme\altaffilmark{1}, M. A. Amo-Baladr\'on\altaffilmark{2}, J. Mart\'{i}n-Pintado\altaffilmark{2}, R. Mauersberger \altaffilmark{3}, L. Bronfman \altaffilmark{4}, S. Mart\'{i}n \altaffilmark{3}}   
\altaffiltext{1}{Instituto de Radioastronom\'{i}a Milim\'etrica (IRAM), Av. Divina Pastora 7, Local 20, E-18012 Granada, Spain}
\altaffiltext{2}{Centro de Astrobiolog\'ia (CSIC/INTA), Ctra. de Torrej\'on a Ajalvir km 4, E-28850, Torrejon de Ardoz, Madrid, Spain}
\altaffiltext{3}{Joint ALMA Observatory, Av. El Golf 40, Piso 18, Las Condes, Santiago de Chile, Chile}
\altaffiltext{4}{Departamento de Astronom\'{i}a, Universidad de Chile, Casilla 36-D, Santiago de Chile, Chile}

\begin{abstract} 
We study the disk-halo interaction, in the context of orbits and Giant Molecular loops (GMLs) in the Galactic center (GC) region. We present a large scale survey in the central kpc in
SiO(J=2-1), HCO$^+$(J=1-0) and H$^{13}$CO$^+$ (J=1-0), observations in 3mm lines
toward a region in two clumps M+5.3-0.3 and M-3.8+0.9 placed in the
foot point of two molecular loops, and observations toward selected
positions, to trace accretion of the gas from the halo.
\end{abstract}

\begin{itemize}
\item A large scale survey of the GC region in the J=2-1 
transition of SiO and the J=1-0 transition of HCO$^+$ and H$^{13}$CO$^+$
(FWHM=3.6') was conducted using the NANTEN telescope to study cloud
conditions, heating mechanisms and chemistry \citep{Riquelme_et_al_2010a}. The observed region
covers an area between $-5^o.75<l<5^o.625$ and $-0^o.6875<b<1^o.35425$. This area includes the
``Central Molecular Zone'' and five molecular clouds, from the ``Peripheral
Molecular Zone'' (PMZ). An enhancement of the SiO/HCO$^+$ line intensity ratio is found at the 
"foot points" of the GMLs \citep{Fukui_et_al_2006} and toward the $1^o.3$ Complex, which
indicates the presence of shocks.
	
\item Higher spatial resolution (FWHM=38'') observations of 3mm lines were performed toward selected regions of two PMZ cloud: M-3.8+0.9 and M+5.3-0.3, using the Mopra telescope. The maps reveal structures at small scales in the SiO emission, an evidence of the presence of shocks
(\citealp{Martin-Pintado_et_al_1992,Martin-Pintado_et_al_1997}). Both mapped clumps show large differences between the spatial distribution of the SiO and the HCO$^+$ emission, which indicates
differences in the chemistry and physical properties within the
clumps. The SiO emission in the M-3.8+0.9 cloud presents narrow profiles (20 km/s) in comparison
with the HCO$^+$ profiles (50 km/s, see Fig. 1), thus, shocked gas is dynamically more confined than the HCO$^+$. Also
remarkable is the decrease of the HNCO emission that we find in the
M+5.3-0.3 cloud as compared with the M-3.8+0.9
cloud. \citet{Martin_et_al_2008} claim that the HNCO molecule is easily photodissociated by UV radiation.
Therefore, it might be possible that stellar winds produced by star clusters in the disk could
play a role in the generation of the shocks  in the molecular cloud 
M+5.3-0.3. 
\begin{figure}
\begin{center}
\hbox{
\includegraphics[width= 2.5in, angle=0]{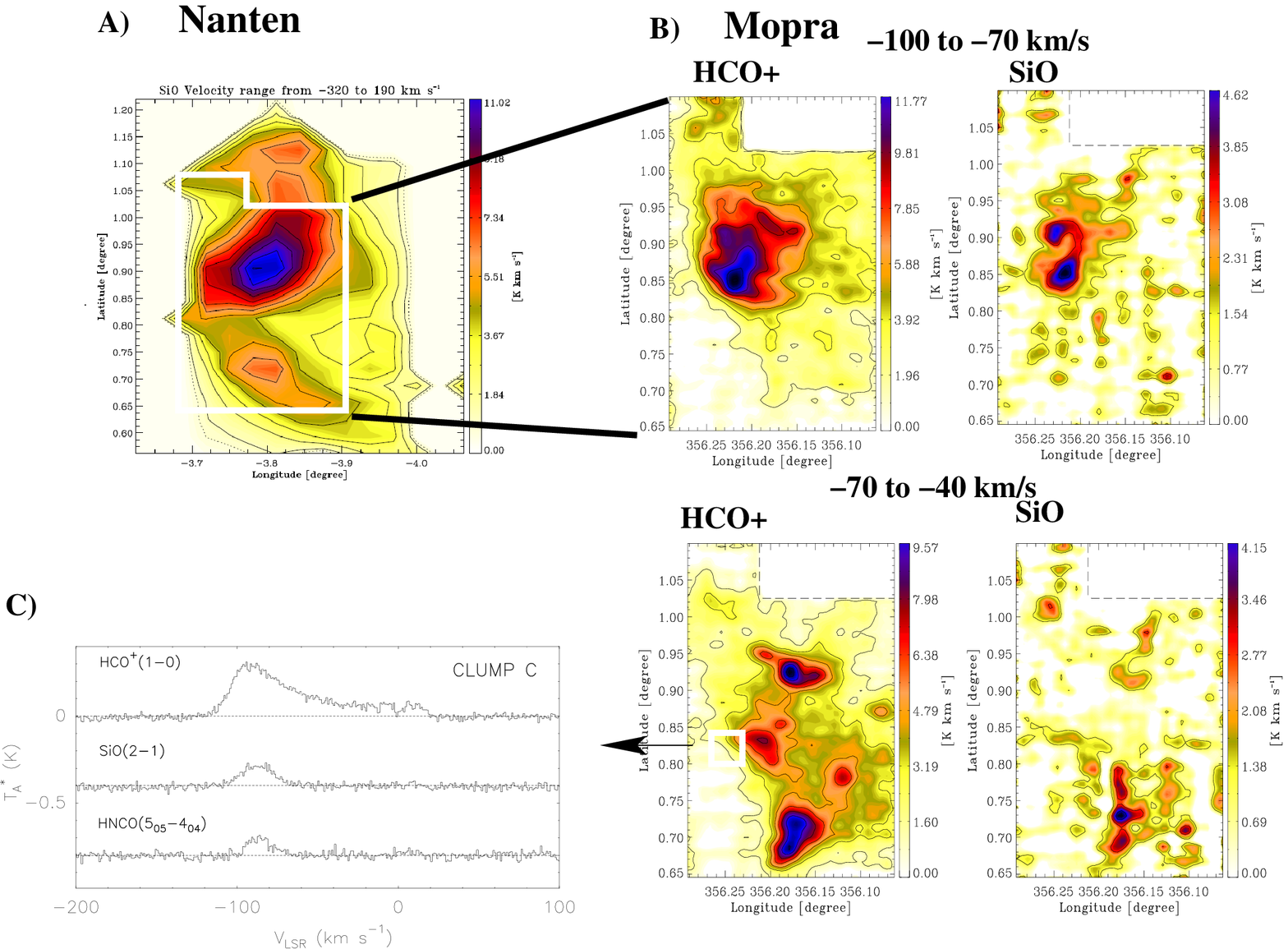}
\includegraphics[width= 2.5in, angle=0]{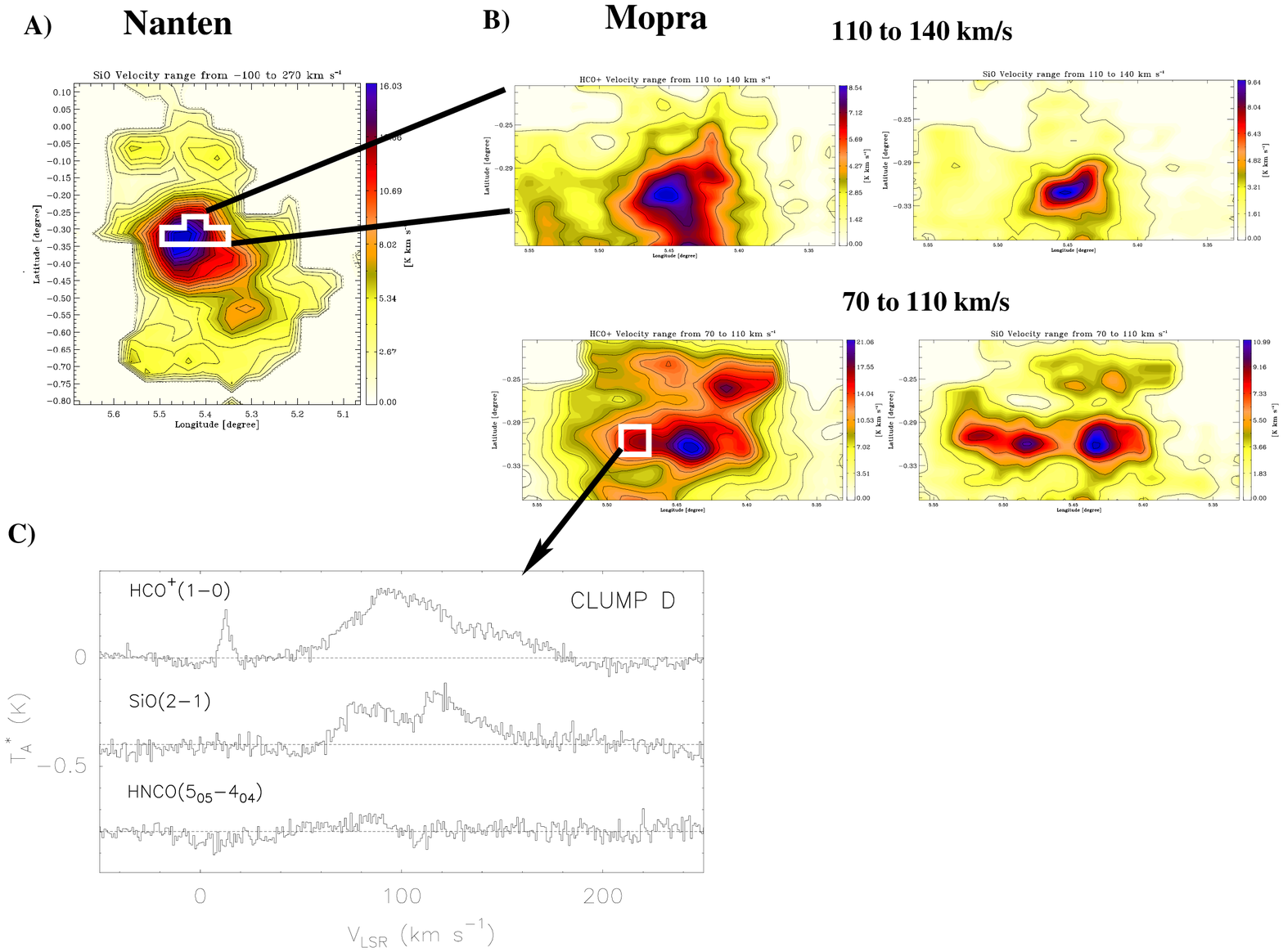}
}
\caption {\scriptsize Left: A) Integrated intensity map of SiO emission in M-3.8+0.9. B) Mopra observations. Top: Integrated intensity map from -110 to -70 km/s in SiO (right) and HCO$^+$ (left). Bottom: Integrated intensity map from -70 to -40 km/s in SiO (right) and HCO$^+$ (left). C) Composite spectra over a region of 1.8$\times$1.2 arcmin in HCO$^+$, SiO, HNCO.
 Right: A) Integrated intensity map of SiO emission of M+5.3-0.3. B) Mopra observation. Top: Integrated intensity map from 70 to 110 km/s in SiO(left) and HCO$^+$(right). Bottom: Integrate intensity map from 110 to 140 km/s in SiO(left) and HCO$^+$(right). C)Composite spectra over a region of 1.9$\times$1.2 arcmin in HCO$^+$, SiO, HNCO.}
\label{cobertura}
\end{center}
\end{figure}

\item The $^{12}$C/$^{13}$C isotopic ratio reflects the relative degree of
primary to secondary processing in stars. $^{12}$C is formed on
rapid time scale in intermediate and high-mass stars, whereas  $^{13}$C
is produced primarily via CNO processing of $^{12}$C seeds
from earlier stellar generation, in a lower time scale in low and
intermediate mass stars or novae. Using the IRAM 30m telescope, we
measure this isotopic ratio in 7 selected positions \citep{Riquelme_et_al_2010b}, 5 toward the disk-halo interaction sites (3 in the GMLs and 2 in
the x$_1$-x$_2$ orbits interaction places \citep{Binney_et_al_1991}), and
2 in the Galactic plane tracing ``standard'' GC gas.
Our preliminary results clearly point to a higher $^{12}$C/$^{13}$C
isotopic ratios (50$-$70) toward the positions and velocity
components associated with disk-halo interaction sites, compared with
the ``standard'' GC values (20$-$25). This result clearly shows that
less-processed gas is being accreted toward the GC at these particular
places.
\end{itemize}


\begin{thebibliography}{}

\bibitem[{{Binney} {et~al.}(1991){Binney}, {Gerhard}, {Stark}, {Bally}, and {Uchida}}]{Binney_et_al_1991} {Binney}, J., {Gerhard}, O.~E., {Stark}, A.~A.,{et~al.}  1991, \mnras, 252, 210

\bibitem[{{Bitran} {et~al.}(1997){Bitran}, {Alvarez}, {Bronfman}, {May}, \&   {Thaddeus}}]{Bitran_et_al_1997} {Bitran}, M., {Alvarez}, H., {Bronfman}, L., {et~al.} 1997, \aaps, 125, 99

\bibitem[{{Fukui} {et~al.}(2006){Fukui}, {Yamamoto}, {Fujishita}, {Kudo}, {Torii}, {Nozawa}, {Takahashi}, {Matsumoto}, {Machida}, {Kawamura},
  {Yonekura}, {Mizuno}, {Onishi}, \& {Mizuno}}]{Fukui_et_al_2006}
{Fukui}, Y.,{Yamamoto}, H., {Fujishita}, M. {et~al.} 2006,
  Science, 314, 106

\bibitem[{{Mart{\'{\i}}n} {et~al.}(2008){Mart{\'{\i}}n}, {Requena-Torres},
  {Mart{\'{\i}}n-Pintado}, \& {Mauersberger}}]{Martin_et_al_2008}
{Mart{\'{\i}}n}, S., {Requena-Torres}, M.~A., {Mart{\'{\i}}n-Pintado}, J., {et~al.} 2008, \apj, 678, 245

\bibitem[{{Mart\'{i}n-Pintado} {et~al.}(1992){Mart\'{i}n-Pintado}, {Bachiller},
  \& {Fuente}}]{Martin-Pintado_et_al_1992}
{Mart\'{i}n-Pintado}, J., {Bachiller}, R., \& {Fuente}, A. 1992, \aap, 254, 315

\bibitem[{{Mart\'{i}n-Pintado} {et~al.}(1997){Mart\'{i}n-Pintado}, {de
  Vicente}, {Fuente}, \& {Planesas}}]{Martin-Pintado_et_al_1997}
{Mart\'{i}n-Pintado}, J., {de Vicente}, P., {Fuente}, A., \& {Planesas}, P.
  1997, \apjl, 482, L45

\bibitem[{{Riquelme} {et~al.}(2010){Riquelme}, {Bronfman}, {Mauersberger}, {May},\& {Wilson}}]{Riquelme_et_al_2010a}
{Riquelme}, D., {Bronfman}, L., {Mauersberger}, R., {May}, J.,\& {Wilson}, T.L. 2010, submitted

\bibitem[{{Riquelme} {et~al.}(2010){Riquelme}, {Amo-Baladr\'on},{Mart\'{i}n-Pintado}, {Mauersberger}, {Mart\'{i}n},  \& {Bronfman}}]{Riquelme_et_al_2010b}
{Riquelme}, D., {Amo-Baladr\'on}, M.A, {Mart\'{i}n-Pintado}, J.,{Mauersberger}, R., {Mart{\'{\i}}n}, S.,\& {Bronfman}, L. 2010, in preparation

\end{thebibliography}
\end{document}